\documentclass[12pt]{iopart}

\usepackage{amssymb}

\RequirePackage{filecontents}
\usepackage{xcolor}
\usepackage{booktabs}
\usepackage{multirow}
\usepackage{epstopdf}
\usepackage{graphicx}

\begin{document}

\newcommand{\bra}{\left\langle}
\newcommand{\ket}{\right\rangle}
\newcommand{\tbox}[1]{\mbox{\tiny #1}}

\title{Statistical properties of mutualistic-competitive random networks}


\author{C. T. Mart\'{\i}nez-Mart\'{\i}nez}
\address{Instituto de F\'isica, Benem\'erita Universidad Aut\'onoma de Puebla,
Apartado Postal J-48, Puebla 72570, Mexico}
\address{Institute for Biocomputation and Physics of Complex Systems (BIFI), University of Zaragoza, 50018 Zaragoza, Spain}

\author{J. A. M\'endez-Berm\'udez}
\address{Instituto de F\'isica, Benem\'erita Universidad Aut\'onoma de Puebla, 
Apartado Postal J-48, Puebla 72570, Mexico}

\author{Thomas Peron}
\address{Instituto de Ci\^{e}ncias Matem\'{a}ticas e de Computa\c{c}\~{a}o, Universidade de S\~{a}o Paulo - S\~{a}o Carlos, SP 13566-590, Brazil}
\ead{thomas.peron@usp.br}

\author{Yamir Moreno}
\address{Institute for Biocomputation and Physics of Complex Systems (BIFI), University of Zaragoza, 50018 Zaragoza, Spain}
\address{Department of Theoretical Physics, University of Zaragoza, 50009 Zaragoza, Spain}
\address{ISI Foundation, Turin, Italy}

\begin{abstract}
Mutualistic networks are used to study the structure and processes inherent to mutualistic relationships. 
In this paper, we introduce a random matrix ensemble (RME) representing the adjacency matrices of 
mutualistic networks composed by two vertex sets of sizes $n$ and $m-n$. Our RME depends on three 
parameters: the network size $n$, the size of the smaller set $m$, and the connectivity between the two
sets $\alpha$, where $\alpha$ is the ratio of current adjacent pairs over the total number of possible 
adjacent pairs between the sets. 
We focus on the the spectral, eigenvector and topological properties of the RME by computing, 
respectively, the ratio of consecutive eigenvalue spacings $r$, the Shannon entropy of the 
eigenvectors $S$, and the Randi\'c index $R$.
First, within a random matrix theory approach (i.e.~a statistical approach), we identify a parameter 
$\xi\equiv\xi(n,m,\alpha)$ that scales the average normalized measures $\left< \overline{X} \right>$ 
(with $X$ representing $r$, $S$ and $R$).
Specifically, we show that 
(i) $\xi\propto \alpha n$ with a weak dependence on $m$, and 
(ii) for $\xi<1/10$ most vertices in the mutualistic network are isolated, while for $\xi>10$ the network 
acquires the properties of a complete network, i.e.,~the transition from isolated vertices to a complete-like
behavior occurs in the interval $1/10<\xi<10$. 
Then, we demonstrate that our statistical approach predicts reasonably well the properties of real-world
mutualistic networks; that is, the universal curves $\left< \overline{X} \right>$ vs.~$\xi$ show good
correspondence with the properties of real-world networks.
\end{abstract}




\section{Introduction}

Many real-world networks can be represented as having two types of nodes grouped into two disjoint sets, such that nodes within the same set are not adjacent.  A network satisfying this definition is denominated a \emph{bipartite network}, and examples of such a structure can be found across a broad range of systems~\cite{newman2018networks}. For instance, there are several types of social networks that have a natural bipartite representation, such as the actor-movie network~\cite{amaral2000classes,newman2001scientific}, where actors are linked to movies in which they were cast; coauthorship networks in which the two sets of nodes are authors and papers, while the edges reveal the authorship of the latter~\cite{newman2018networks,newman2001scientific}; and networks linking people to the social events they attended~\cite{doreian2004generalized}. Other noteworthy examples include recommendation systems~\cite{zhou2007bipartite}, networks of heterosexual contacts~\cite{liljeros2001web}, among others~\cite{newman2018networks}. Bipartite networks are particularly relevant in ecology~\cite{bascompte2013mutualistic}, for they naturally encode the structure of mutualistic interactions of plant-pollinator, seed dispersal, and host-parasite networks~\cite{bascompte2013mutualistic,pavlopoulos2018bipartite}. In the theoretical domain, bipartite networks are useful to encapsulate the structure of networks formed by distributions of subgraphs~\cite{karrer2010random}, where one group is formed by nodes, and the other group by subgraphs to which the nodes are attached. Bipartite networks can also offer an alternative representation of hypergraphs by mapping nodes and hyperedges into disjoint groups, where an original hyperedge is connected by an edge to the nodes it encompasses in the original hypergraph representation~\cite{newman2018networks,battiston2020networks}.      

In all the above examples, the disjoint sets aggregate nodes that have the same type, and the edges only run from one set to the other. However, from a bipartite network it is also possible to unfold relationships pertaining to nodes that belong to the same group. For example, from the actor-movie network we are able to construct another network revealing who acted with whom: if two actors are connected to at least one common movie in the original bipartite network, a link is then created in a new network informing that they have collaborated at least once. The new generated network is oftentimes referred to as the \emph{one-mode projection} of the bipartite structure~\cite{newman2018networks}, owing to the fact that all nodes have the same type in the new mapping. The same procedure can be applied in order to obtain the one-mode projection of the movie set, thereby connecting any two movies that have at least one actor in common. For any bipartite network, there are thus two one-mode projections, one associated with each type of node.   

One-mode projections are useful, since they allow the investigation of aspects that might be hidden or simply not apparent in the original bipartite network. In certain applications, however, neither the bipartite network nor its two one-mode projections alone are sufficient to accurately model the dynamics of the system that they represent. For example, the population dynamics of plant-pollinator networks depend crucially on both mutualistic interactions and intra-group competitions~\cite{bascompte2013mutualistic}. Mutualistic connections in such an ecological community can be empirically mapped by field observations; that is, the information that a given plant is pollinated by a given animal species can be stored in a bipartite network where the two types of nodes are plants and pollinators~\cite{bascompte2013mutualistic}. Intra-group
connections, which quantify how strongly plants and pollinators 
compete among themselves for resources, are, on the other hand, not readily accessed and need to be inferred via one-mode projections onto 
the pollinator and plant groups~\cite{bascompte2013mutualistic}. There is not a single way to project 
the bipartite networks in order to obtain the intra-group connections in this context. Traditional dynamical models adopt a mean-field description by treating the intra-groups to be fully connected, creating then a scenario in which all plants and pollinators compete equally for resources~\cite{bastolla2009architecture}. This has been argued to be a very strong assumption since it neglects completely the rich structure of the bipartite mutualistic interactions observed in real ecological communities~\cite{gracia2018joint}. To overcome this limitation, heterogeneous competition schemes have been recently introduced~\cite{gracia2018joint} and consist of projecting the bipartite connections as described above for the actor-movie network: if two given pollinators (plants) share at least one common plant (pollinator), an edge is created between them representing their competitive interaction. In both competition scenarios, homogeneous and heterogeneous, the network underlying the interactions
of the dynamical model is not the original bipartite structure, but rather the union of the latter with       
its one-mode projections.

While the structure and the dynamics of pure bipartite networks have been scrutinized over the past years (see, e.g.,~\cite{newman2018networks,bascompte2013mutualistic}), little is known about the statistics of spectral properties of the networks that are created by the union of the bipartite connections with their one-mode projections. For this reason, and given the fact the network spectra is intrinsically related with the stability of dynamical processes, in this paper we put forward a thorough characterization of the spectra of such networks, bearing in mind possible implications to the dynamics of real plant-pollinator networks. More specifically, here we characterize the statistics of the eigenvalues of matrices generated by grouping the adjacency matrix of random bipartite networks with the matrices obtained from their respective two one-mode projections (see Fig.~\ref{Fig1} for an illustration). Our interest in this network representation stems from the fact that it mimics the structure of Jacobian matrices of dynamical models describing plant-pollinator communities more closely than traditional random matrix ensembles~\cite{bastolla2009architecture,gracia2018joint}. 

We employ three measurements to assess the statistical regimes of mutualistic-competitive random networks as a function of size and connectivity, namely, the ratio between consecutive eigenvalue spacings, the Shannon entropy related to the eigenvectors of the adjacency matrix, and the Randi\'c index. Very recently, these quantities have been successfully applied to the characterization of other network ensembles (see, e.g., Refs.~\cite{PRR20,MMRS20} and references therein).  Our scaling analysis reveals that the three measurements exhibit a universal behavior as a function of the average degree. The obtained universal behavior highlights three markedly different statistical regimes: at sufficiently low and high connectivity, the spectral and eigenvector statistics of the networks coincides with those of the Poisson Ensemble (PE) and the Gaussian Orthogonal Ensemble (GOE) of Random Matrix Theory (RMT)~\cite{metha}, respectively; for intermediate connectivity, the networks undergo a delocalization-to-localization transition that mediates the latter regimes. We further show, unexpectedly, that real--world ecological networks follow with a reasonably good agreement the universal behavior reported for the random networks -- a result that, as we argue, indicates that such ecological communities might operate in a regime of maximal complexity.  

The remainder of this paper is organized as follows: In Section~\ref{model} we introduce the random network ensemble we study and the 
measurements used to characterize the network properties. Subsequently, in Section~\ref{universality} we discuss the scaling and 
the universality properties of the random networks with projected edges. We then apply the scaling approach to a set of real plant-pollinator networks in Section~\ref{real}. Section~\ref{conclusions} is dedicated to our conclusion and perspectives for future works.

\section{Network model and measures}
\label{model}

\subsection{Random mutualistic-competitive networks}

\begin{figure}[t]
\centering
\includegraphics[width=0.3\textwidth]{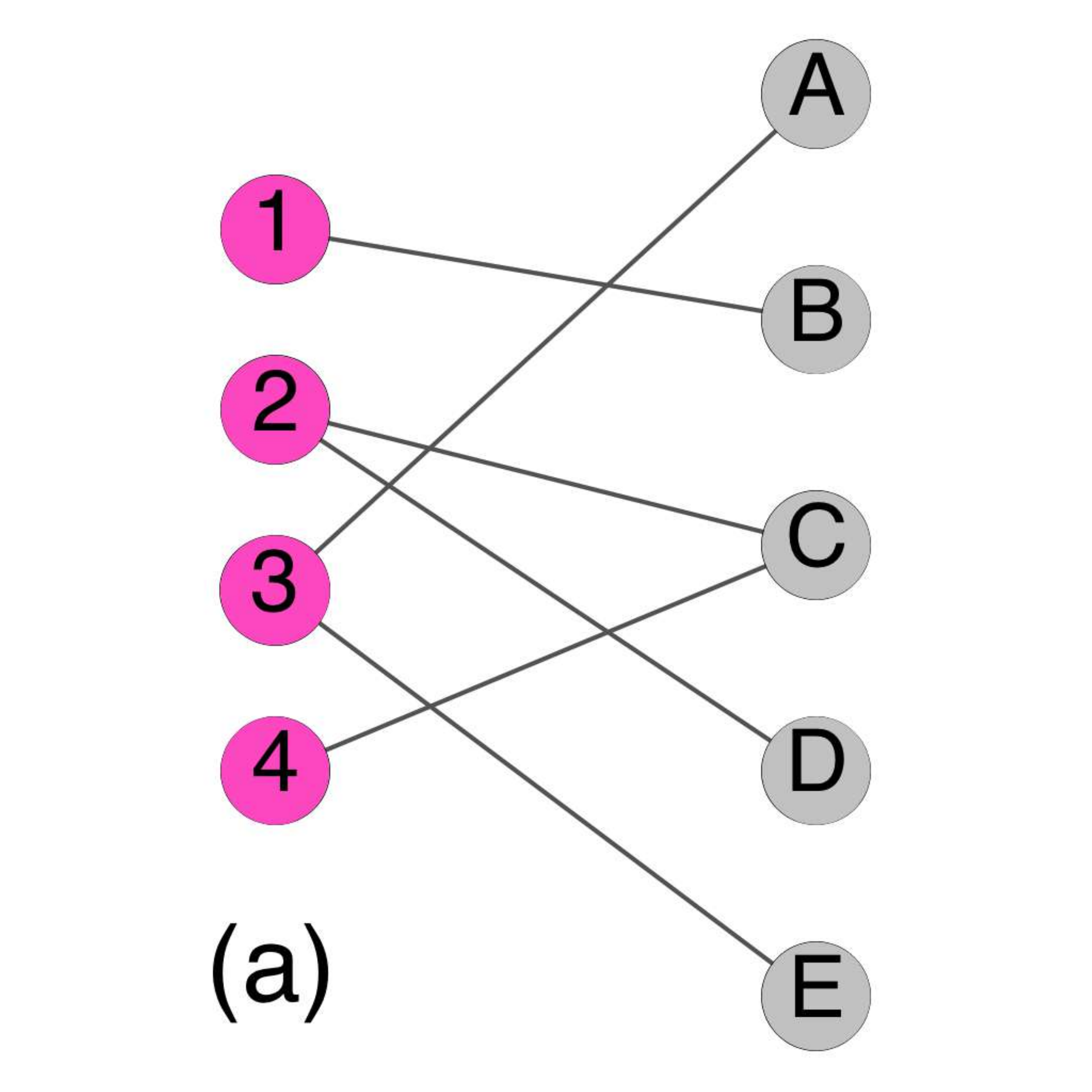}
\includegraphics[width=0.3\textwidth]{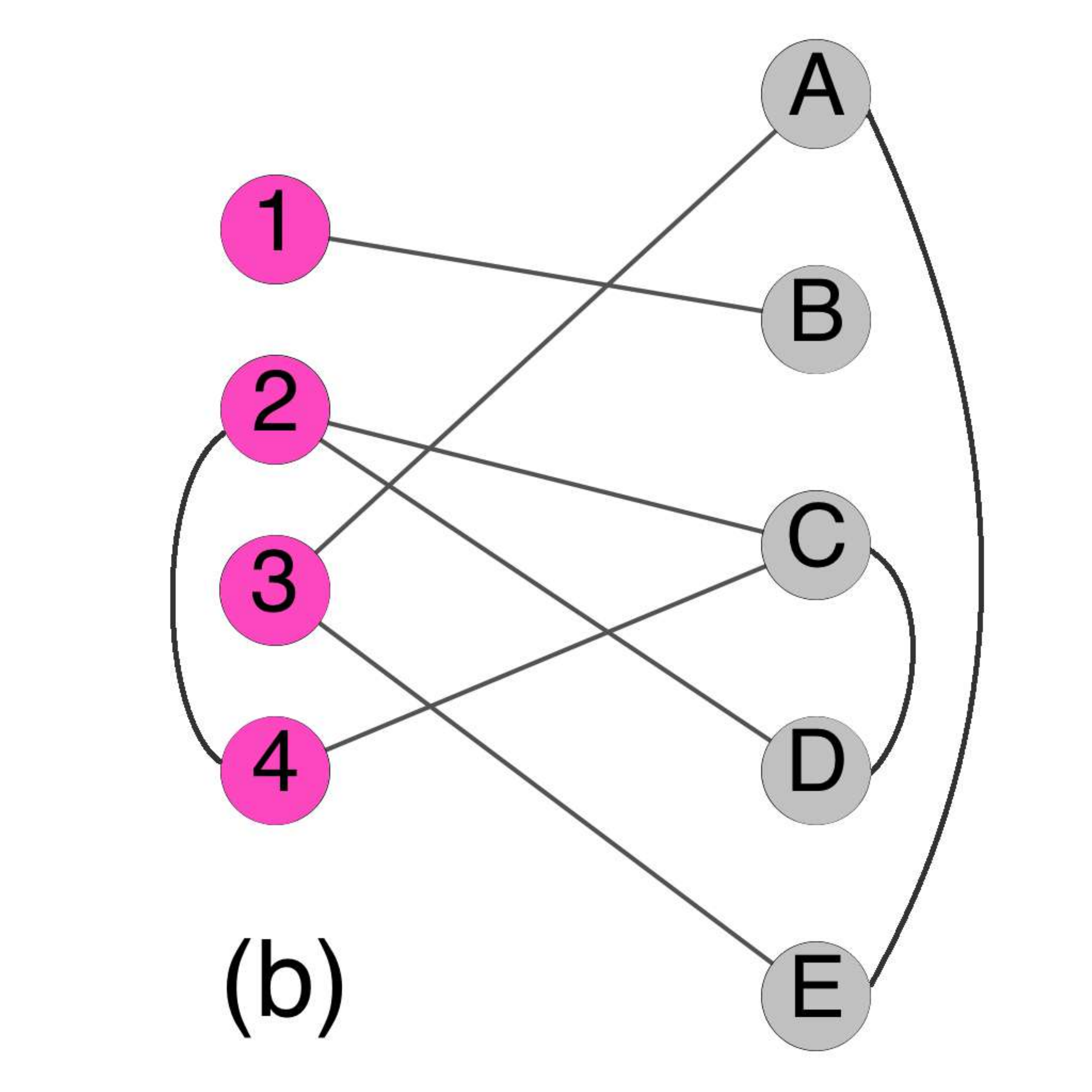}
\caption{Example of (a) bipartite network representing the mutualistic interactions between two groups for $n = 9$, $m = 4$ and $\alpha = 0.35$, and 
(b) the same mutualistic network with its competition edges projected onto each group.}
\label{Fig1}
\end{figure}

We start with a bipartite network composed by two disjoint sets with $m$ and $n-m$ vertices each such that 
there are no adjacent vertices within the same set, being $n$ the total number of vertices in the bipartite 
network. The connectivity between both sets is quantified by the parameter $\alpha$ which is the ratio of 
current adjacent pairs over the total number of possible adjacent pairs; that is, vertices are isolated
when $\alpha=0$, whereas the bipartite graph is complete for $\alpha=1$. Vertices are connected randomly. 
An example of a bipartite network with $n = 9$, $m = 4$ and $\alpha = 0.35$ is shown in Fig.~\ref{Fig1}(a).
Then, the mutualistic-competitive network is constructed by establishing connections between elements of the same set 
when they are connected to a common vertex of the other set. The mutualistic-competitive network corresponding to
the bipartite network of Fig.~\ref{Fig1}(a) is presented in Fig.~\ref{Fig1}(b).

Here we follow a recently introduced approach under which the adjacency matrices of random 
graphs and networks are represented by RMT ensembles; see the application of this approach on 
Erd\"os-R\'{e}nyi graphs~\cite{MAM15}, random rectangular graphs~\cite{AMGM18}, 
$\beta$-skeleton graphs~\cite{AME19}, multiplex and multilayer networks~\cite{MFMR17}, and 
bipartite networks~\cite{MAMPS19}.
Accordingly, we define the elements of the $n\times n$ adjacency matrix $\mathbf{A}$ of a mutualistic 
network as
\begin{equation}
A_{ij}=\left\{
\begin{array}{cl}
\sqrt{2} \epsilon_{ii} \ & \mbox{for $i=j$}, \\
\epsilon_{ij} & \mbox{if there is an edge between vertices $i$ and $j$},\\
0 \ & \mbox{otherwise}.
\end{array}
\right.
\label{Aij}
\end{equation}
We choose $\epsilon_{ij}$ as statistically-independent random variables drawn from a normal 
distribution with zero mean and unity variance. Also, $\epsilon_{ij}=\epsilon_{ji}$, since the network is 
assumed as undirected. 
Indeed, according to definition (\ref{Aij}), diagonal random matrices are obtained for $\alpha=0$ (known 
as PE in RMT~\cite{metha}), whereas the GOE (i.e.,~full real and 
symmetric random matrices~\cite{metha}) is recovered when $\alpha=1$. Therefore, a transition from the PE to the 
GOE should be observed by increasing $\alpha$ from zero to one, for any given pair $(n,m)$.

\begin{figure}[t]
\centering
\includegraphics[width=0.9\textwidth]{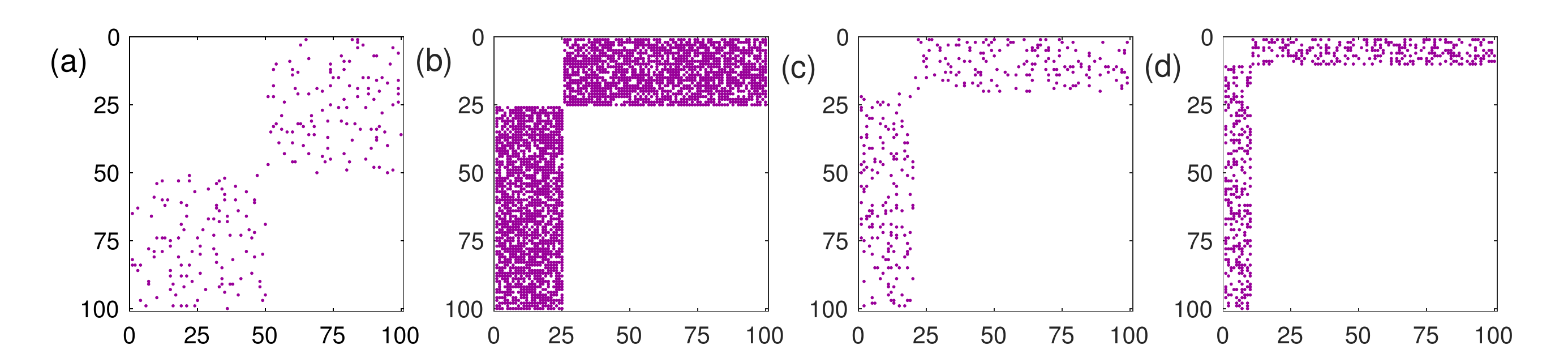}
\caption{Nonzero adjacency matrix elements of bipartite random networks for some combinations of 
$m/n$ and $\alpha$: (a) $m/n=1/2$ and $\alpha =0.05$, (b) $m/n=1/4$ and $\alpha =0.75$, (c) $m/n=1/5$ 
and $\alpha =0.1$, (d) $m/n=1/10$ and $\alpha =0.3$. In all cases $n=100$.}
\label{Fig2}
\end{figure}
\begin{figure}[t]
\centering
\includegraphics[width=0.9\textwidth]{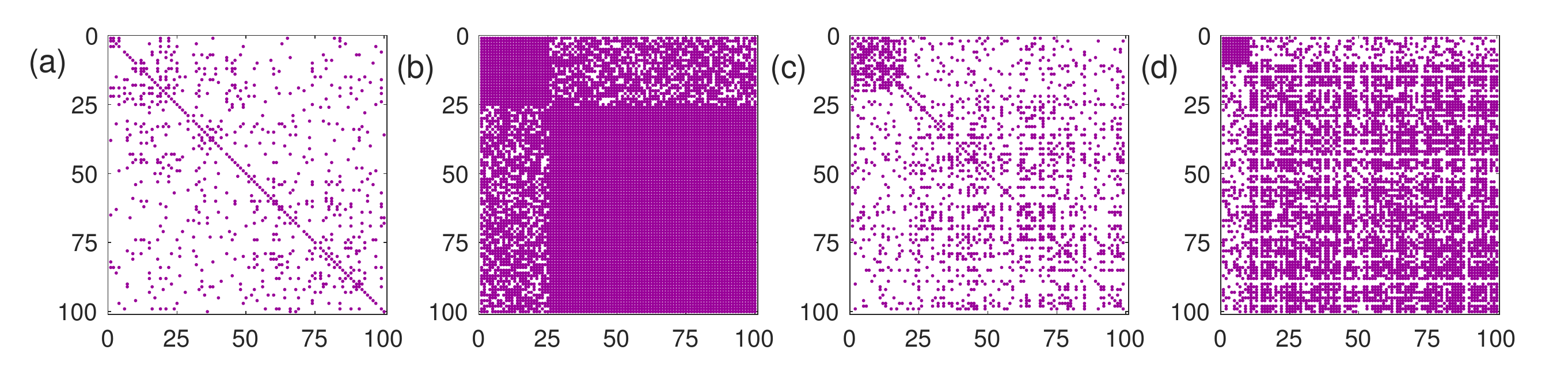}
\caption{Nonzero adjacency matrix elements of the mutualistic-competitive random networks corresponding to the
bipartite networks of Fig.~\ref{Fig2}.}
\label{Fig3}
\end{figure}

In Fig.~\ref{Fig2}, we show examples of adjacency matrices of random bipartite networks with $n=100$ 
vertices and some combinations of $m$ and $\alpha$; while in Fig.~\ref{Fig3} we present the adjacency 
matrices of the corresponding mutualistic-competitive networks. Note that when labeling the vertices according to the 
set they belong to, the adjacency matrices of both bipartite and mutualistic-competitive networks have a $2\times 2$ 
block structure.
Notice also that, in contrast to bipartite networks, since connections between vertices of the same set 
are allowed in mutualistic-competitive networks, the diagonal blocks of the corresponding adjacency matrices are not 
null matrices.

Below we define $m$ (resp. $n-m$) as the number of vertices of the smaller (bigger) set. In this respect, 
the case $m=n/2$ is a limiting case where both sets have the same number of vertices, $m=n-m$. 
Moreover, the case $m=1$ is another limiting case in which the smaller set consists of a single vertex.
Thus, in what follows we will consider random mutualistic-competitive networks characterized by the parameter set 
$(n,m,\alpha)$ with $1\le m\le n/2$ and $0\le \alpha \le 1$. 
Notice that the case $m>n/2$ is redundant because it is equivalent to the interchange of the sets.

\subsection{Spectral and topological measures}

We characterize the spectral and eigenvector properties of the 
randomly-weighted adjacency matrices of mutualistic-competitive networks by the use of two well-known RMT 
measures: the ratio between consecutive eigenvalue spacings $r$~\cite{ABG13} and the information 
or Shannon entropies $S$~\cite{MK98}, whereas to probe topological properties we use the Randi\'c index 
$R$~\cite{R}, one of the best studied topological indices in mathematical chemistry.

On the one hand, given the ordered spectra $\{ \lambda_i \}$ ($i=1,\ldots,n$) and the corresponding 
normalized eigenvectors $\Psi^i$, i.e., $\sum_{j=1}^n | \Psi^i_j |^2 =1$, the ratio $r_i$ 
and the entropy $S_i $ are given by~\cite{ABG13,MK98}
\begin{equation}
\label{r}
r_i = \frac{\min(\lambda_{i+1}- \lambda_i,\lambda_{i}- \lambda_{i-1})}{\max(\lambda_{i+1}- \lambda_i,\lambda_{i}- \lambda_{i-1})} 
\end{equation}
and
\begin{equation}
\label{S}
S_i = -\sum_{j=1}^n \left| \Psi^i_j \right|^2 \ln \left| \Psi^i_j \right| ^2 \ ,
\end{equation}
respectively.

It is pertinent to mention that $S$, which quantifies the extension of eigenvectors in a given basis, has 
been widely used to study the localization characteristics of the eigenvectors of random graphs and 
network models. 
Among the vast amount of studies available in the literature, we can mention (as relevant examples to
the present study) that $S$ was used to 
find the universal parameters able to scale the eigenvector properties of multiplex and multilayer 
networks~\cite{MFMR17} and bipartite graphs~\cite{MAMPS19}.
In contrast, $r$ has been scarcely used in studies of networks; for a recent exception see 
Ref.~\cite{TFM20}, were $P(r)$ served to characterize the percolation transition 
in weighted Erd\"os-R\'{e}nyi graphs.
We believe that the lack of use of $r$ in network studies is mainly due to the fact that the introduction
of $r$ is relatively recent. In fact, most studies of spectral properties of random graphs and networks, 
from a RMT point of view, are based on the nearest-neighbor energy level spacing distribution $P(s)$, 
see e.g.~\cite{MAM15} and the references therein.
However, here we prefer to use $\left< r \right>$, instead of $P(s)$, because 
the calculation of the ratios $r_i\equiv \min(s_i, s_{i+1})/\max(s_i, s_{i+1})$ (with 
$s_i=(\lambda_{i+1}-\lambda_i)/\Delta$, $\Delta$ being the mean eigenvalue spacing) do not 
require the spectrum unfolding~\cite{metha}, a task that may become cumbersome mainly for the spectra of 
real-world systems. Moreover, the spectrum 
unfolding fixes $\left< s \right>=1$ and forbids the use of $\left< s \right>$ as a complexity indicator;
a restriction not applicable to $\left< r \right>$. See e.g.~Ref.~\cite{PRR20} where $\left< r \right>$
has been recently used as a complexity indicator for directed random netwroks.

On the other hand, given a simple connected network with edge set $E(G)$, the Randi\'c connectivity 
index is defined as~\cite{R}
\begin{equation}
\label{R}
R = \sum_{uv\in E(G)} \frac1{\sqrt{d_u d_v}} \ ,
\end{equation}
where $uv$ denotes the edge connecting the vertices $u$ and $v$, and $d_i$ is the degree of the 
vertex $i$. We want to note that the statistical study of $R$ we perform here is justified by the RMT approach
to mutualistic-competitive networks.
This statistical approach, well known in RMT studies, is not widespread in 
studies of topological indices, mainly because topological indices are not commonly applied to 
random graphs and networks; for recent exceptions see~\cite{MMRS20,MMRS21} where average
topological indices have been used as complexity indicators equivalent to traditional RMT measures.
We also notice that the random weights we impose to the adjacency matrix $\mathbf{A}$, as defined 
in Eq.~(\ref{Aij}), do not play any role in the computation of vertex-degree-based indices.

From definitions~(\ref{r}-\ref{R}), when $\alpha=0$ (i.e.,~when all vertices of the mutualistic-competitive network
are isolated) we have $\left< r \right>_{\tbox{PE}}\approx 0.3863$~\cite{ABG13}, 
$\left< S \right>_{\tbox{PE}}=0$ and $\left< R \right>_{\tbox{PE}}=0$.
While when $\alpha=1$ (i.e.~when the mutualistic-competitive network is complete), 
$\left< r \right>_{\tbox{GOE}}\approx 0.5359$~\cite{ABG13},
$\left< S \right>_{\tbox{GOE}}\approx \ln (n/2.07)$~\cite{MK98} and 
$\left< R \right>_{\tbox{GOE}}=n/2$. 
Here and below $\left< \cdot \right>$ denotes the average 
over all eigenvalues/eigenvectors/matrices of an ensemble of mutualistic-competitive networks. 
We just want to add that the values of $\left< r \right>$ reported above for the PE and the GOE 
limits are valid in the large--network--size limit only, see \ref{app1} for a small--network--size analysis
of $\left< r \right>$ at $\alpha=0$ and 1.

\section{Scaling and universality}
\label{universality}

We now apply a scaling approach that has been successfully used to find universal properties of 
random graphs and network models, see e.g.~\cite{MAM15,AMGM18,AME19,MFMR17,MAMPS19}.
We can summarize this approach in the following steps:
(i) plot the average spectral or topological measure $\left< X \right>$ as a function of the parameter $x$, 
which drives the network model from the PE to the GOE regimes, so that both limits can be well identified; 
(ii) normalize the average measure $\left< X \right>$ such that $\left< \overline{X} \right>_{\tbox{PE}}=0$ 
and $\left< \overline{X} \right>_{\tbox{GOE}}=1$;
(iii) define the PE--to--GOE transition point $x^*$ as the value of $x$ such that 
$\left< \overline{X} \right> \approx {\cal C}$ with ${\cal C}\in(0,1)$;
(iv) define the scaling parameter $\xi$ as the ratio $x/x^*$.
Thus, the curves $\left< \overline{X} \right>$ vs.~$\xi$ should fall one on top of the other; that is,
$\left< \overline{X} \right>$ vs.~$\xi$ is a universal curve characterized by the scaling parameter 
$\xi$, where $\xi$ can be explicitly written in terms of the network model parameters.
Therefore, once the universal curve is found, it is possible to identify the network parameters
setting the network properties on the PE and GOE regimes.

\begin{figure}[t]
\centering
\includegraphics[width=0.95\columnwidth]{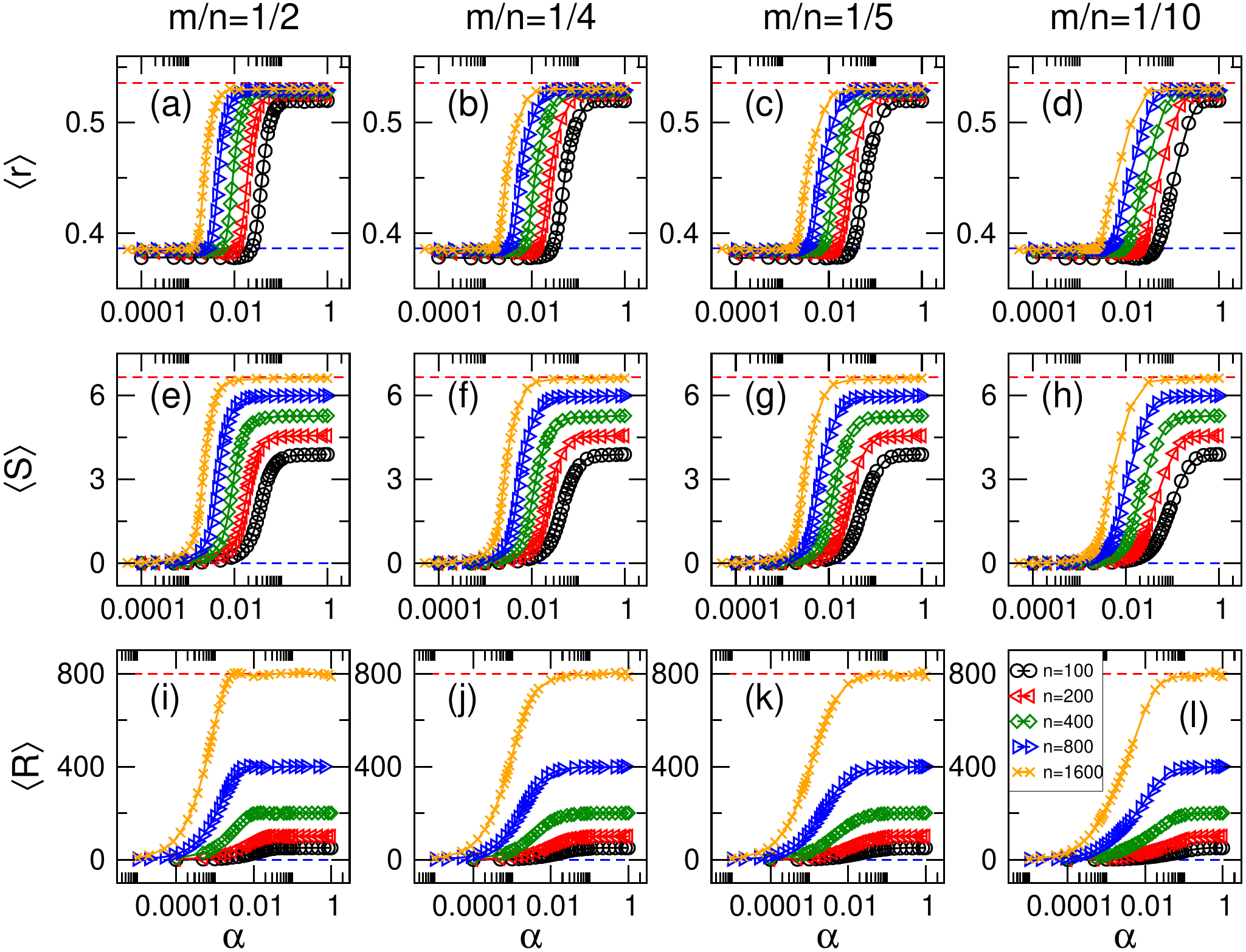}
\caption{(a-d) Average ratio between consecutive eigenvalue spacings $\bra r \ket$, (e-h) average 
Shannon entropy $\bra S \ket$ and (i-l) average Randi\'c index $\bra R \ket$ as a function of the 
connectivity $\alpha$ for random 
mutualistic-competitive networks of size $n$. Four values of the ratio $m/n$ are considered:
(a,e,i) 1/2, (b,f,j) 1/4, (c,g,k) 1/5 and (d,h,l) 1/10. The (blue) red horizontal dashed lines correspond
to the RMT predictions for the (PE) GOE with $n=1600$.
Each symbol was computed by averaging over $10^6/n$ random networks.}
\label{Fig4}
\end{figure}
\begin{figure}[t]
\centering
\includegraphics[width=0.95\columnwidth]{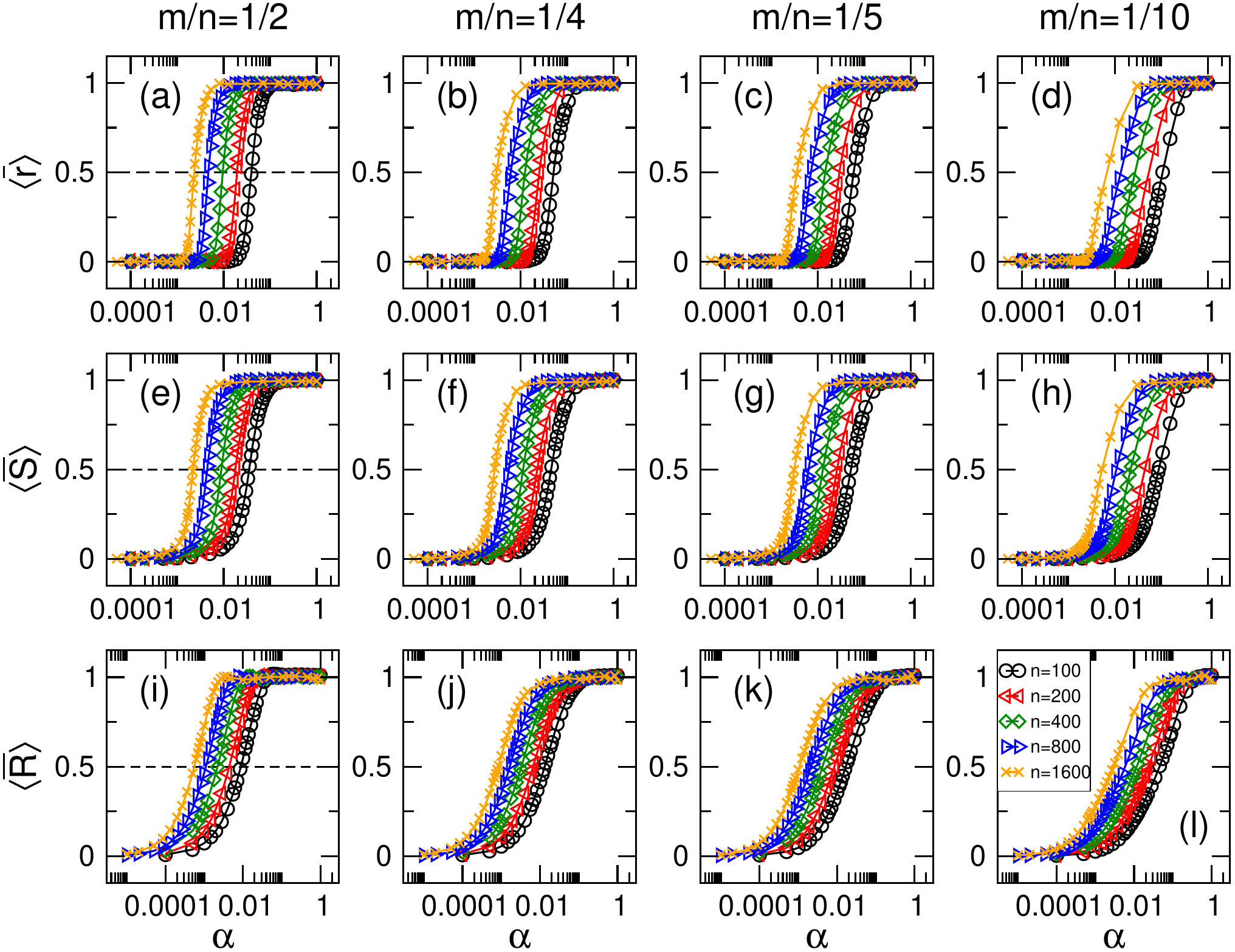}
\caption{Normalized measures (a-d) $\left< \overline{r} \right>$, (e-h) $\left< \overline{S} \right>$ 
and (i-l) $\left< \overline{R} \right>$ as a function of the connectivity $\alpha$ for random 
mutualistic-competitive networks of size $n$. Same curves as in Fig.~\ref{Fig4}. Horizontal dashed lines in
left panels indicate $\bra \overline{X} \ket= 0.5$.}
\label{Fig5}
\end{figure}

Following the steps listed above, in Fig.~\ref{Fig4} we present the average ratio $\bra r \ket$
(upper panels), the average Shannon entropy $\bra S \ket$ (middle panels) and the average Randi\'c 
index $\bra R \ket$ (lower panels) as a function of the connectivity $\alpha$ for mutualistic-competitive random 
networks characterized by different values of $m/n$. Each panel reports five network sizes ranging 
from $n=100$ to 1600. From this figure, it is clear that all curves $\bra X \ket$ vs.~$\alpha$ show the 
transition from the PE to the GOE  (here and below $X$ represents the three measures reported in 
this work: $r$, $S$ and $R$).

\begin{figure}[t]
\centering
\includegraphics[width=0.95\columnwidth]{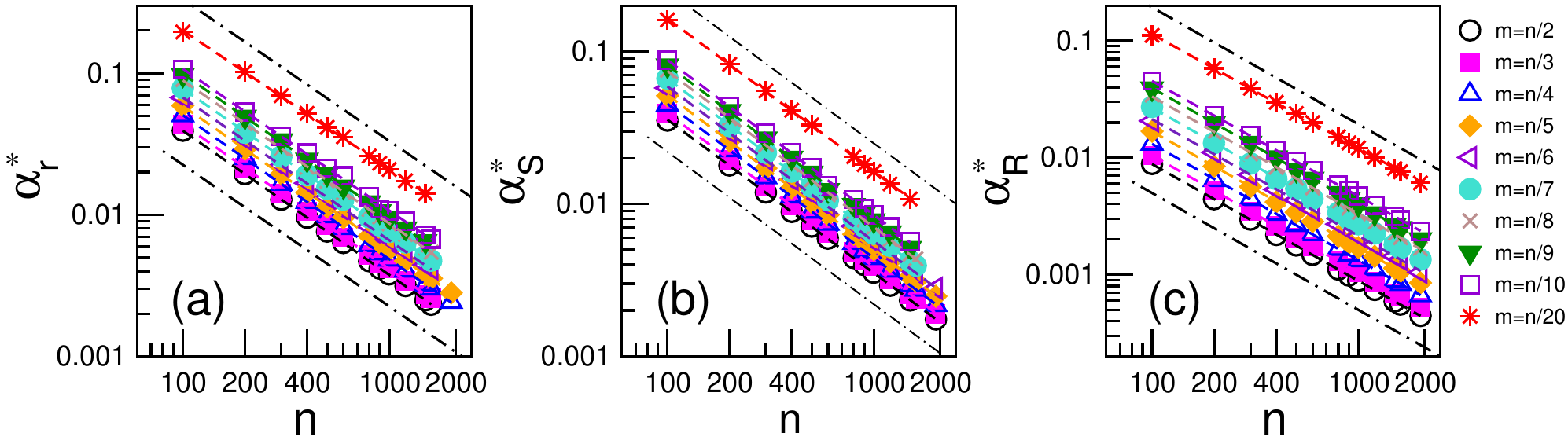}
\caption{Localization--to--delocalization transition point $\alpha^*$ as a function of $n$ for several 
values of the ratio $m/n$. $\alpha^*$ was extracted from curves (a) $\bra \overline{r} \ket$ vs.~$\alpha$, 
(b) $\bra \overline{S} \ket$ vs.~$\alpha$ and (c) $\bra \overline{R} \ket$ vs.~$\alpha$. The dashed lines 
are fittings to the data with Eq.~(\ref{scaling}); the values of $\delta$ and $\lambda$ obtained from these 
fittings are reported in Table~\ref{Table1} and Fig.~\ref{Fig9}, respectively. Dot-dashed lines, shown to 
guide the eye, are proportional to $n^{-1}$.}
\label{Fig6}
\end{figure}

\begin{figure}[h!!]
\centering
\includegraphics[width=0.95\columnwidth]{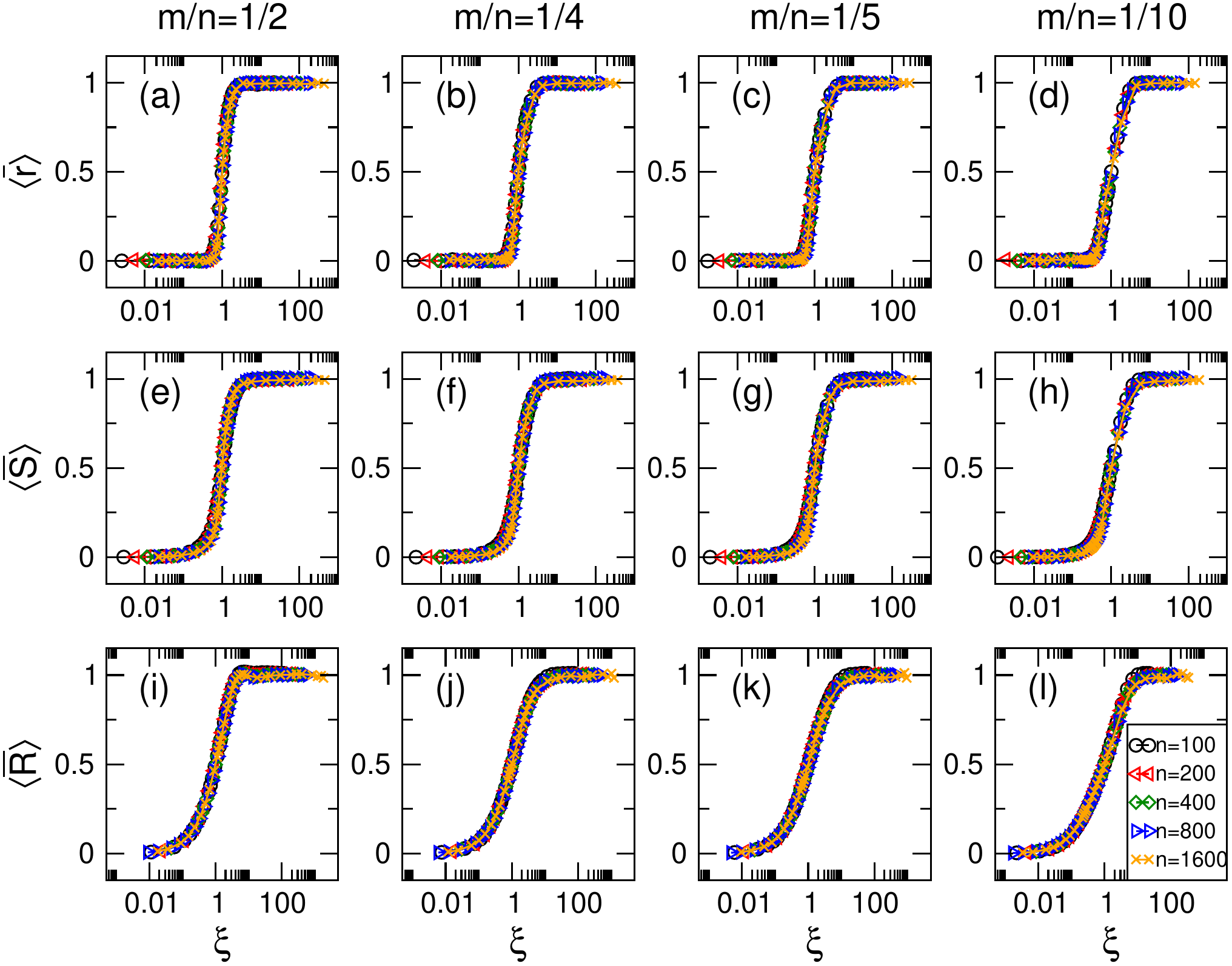}
\caption{ Normalized measures (a-d) $\left< \overline{r} \right>$, (e-h) $\left< \overline{S} \right>$ 
and (i-l) $\left< \overline{R} \right>$ as a function of the scaling parameter $\xi$ for random 
mutualistic-competitive networks of size $n$. Same curves of Fig.~\ref{Fig5}.}
\label{Fig7}
\end{figure}

Then, in Fig.~\ref{Fig5}, we plot again the curves of Fig.~\ref{Fig4} but normalizing $\bra X \ket$ such
that $\left< \overline{X} \right>_{\tbox{PE}}=0$ and $\left< \overline{X} \right>_{\tbox{GOE}}=1$. That 
is, $\left< \overline{r} \right>\equiv[\left< r \right>-\left< r \right>_{\tbox{PE}}]/[\left< r \right>_{\tbox{GOE}}-
\left< r \right>_{\tbox{PE}}]$,
$\left< \overline{S} \right>\equiv\left< S \right>/\left< S \right>_{\tbox{GOE}}$ and
$\left< \overline{R} \right>\equiv\left< R \right>/\left< R \right>_{\tbox{GOE}}$.
We note that while we use $\left< S \right>_{\tbox{GOE}}\approx\ln (n/2.07)$ and 
$\left< R \right>_{\tbox{GOE}}=n/2$, due to small--network--size effects, the values of 
$\left< r \right>_{\tbox{PE}}$ and $\left< r \right>_{\tbox{GOE}}$ are computed numerically; see \ref{app1}. 
Figure~\ref{Fig5} shows that the net effect of increasing the network size $n$ is the
displacement of the curves $\bra \overline{X} \ket$ to the left on the $\alpha$-axis.
Moreover, the fact that the curves are displaced the same amount (in log scale) when doubling $n$ is 
a signature of the scaling of $\bra \overline{X} \ket$ with $n$.
Thus, in order to look for the corresponding scaling parameter we characterize the position of the 
curves $\bra \overline{X} \ket$ vs.~$\alpha$ by extracting the localization--to--delocalization transition 
point $\alpha^*$ that we define as the value of $\alpha$ for which $\bra \overline{X} \ket\approx 0.5$;
i.e.,~the value of $\alpha$ such that $\bra \overline{X} \ket$ is at half of the transition between the 
PE and the GOE. 
In Fig.~\ref{Fig6}, we report the localization--to--delocalization transition point $\alpha^*$ as a function 
of $n$ for several values of the ratio $m/n$. Indeed, the linear trend of the data (in log-log scale) in 
Fig.~\ref{Fig6} implies a power-law relation of the form
\begin{equation}
\label{scaling}
\alpha^* = \lambda n^\delta \ .
\end{equation}
As can be observed in Fig.~\ref{Fig6} (see the dashed lines), Eq.~(\ref{scaling}) provides very good 
fittings to the data.
The values of the power $\delta$ obtained from the fittings in Fig.~\ref{Fig6} (which are reported in 
Table~\ref{Table1}) allowed us to conclude that $\delta\approx 1$ for all the combinations of $(n,m)$
considered here. So, we write
\begin{equation}
\label{xi}
\xi = \frac{\alpha}{\alpha^*} \propto \alpha n \ .
\end{equation}
Therefore, by plotting again the curves $\bra \overline{X} \ket$ now as a function of $\xi$ we observe that 
curves for different mutualistic network sizes $n$ collapse on top of a single curve, see Fig.~\ref{Fig7}.
That is, for a given ratio $m/n$, $\xi$ fixes the spectral and topological properties 
of our randomly-weighted mutualistic-competitive networks.

\begin{table}[h!]
\begin{center}
\scalebox{0.8}{
\begin{tabular}{|c | c |c |c |}
\hline
\hline
& \multicolumn{3}{ |c |}{measure} \\ 
\hline
m/n & $\bra \overline{r} \ket$  & $\bra \overline{S} \ket$ & $\bra \overline{R} \ket$ \\ \hline
1/2 & 0.979 & 1.006 & 0.998  \\
1/3 & 1.001 & 1.009 & 1.001  \\
1/4 & 1.011 & 1.012 & 0.994  \\
1/5 & 1.011 & 1.014 & 0.996  \\
1/6 & 1.007 & 1.015 & 0.998  \\
1/7 & 1.009 & 1.019 & 0.998  \\
1/8 & 1.015 & 1.025 & 1.003  \\
1/9 & 1.017 & 1.022 & 0.997  \\
1/10 & 1.022 & 1.018 & 0.987  \\
1/20 & 1.018 & 1.003 & 0.976  \\ \hline
\hline
\end{tabular}}
\caption{Values of the power $\delta$ obtained from the fittings, with Eq.~(\ref{scaling}), of the 
curves $\alpha^*(X)$ vs.~$n$ of Fig.~\ref{Fig6}.}
\label{Table1}
\end{center}
\end{table}

It is important to add that even though we were able to scale the average ratio between consecutive 
eigenvalue spacings, the average Shannon entropy and the average Randi\'c index of random 
mutualistic-competitive networks, as shown in Fig.~\ref{Fig7}, there is still a dependence (weak, though) of the 
curves $\bra \overline{X} \ket$ vs.~$\xi$ on the ratio $m/n$. Indeed, a similar dependence was 
reported for random bipartite networks in Ref.~\cite{MAMPS19}.
To illustrate this weak dependence, in Fig.~\ref{F10x} we report curves $\bra \overline{X} \ket$ vs.~$\xi$ 
for several values of $m/n$. Here, we can observe that the larger the ratio $m/n$, the sharper the 
PE--to--GOE transition. However, it is relevant to add that our interest is focused on large values of
$m/n$ since we have observed that most real--world mutualistic networks are characterized by ratios
in the interval $(1/3,1/2)$; see the next Section.

From Fig.~\ref{Fig7} we can conclude that the average properties of the random mutualistic-competitive 
network model studied here coincide with those of the PE and the GOE when $\xi<1/10$ and $\xi>10$, 
respectively; while a PE--to--GOE transition regime approximately appears for $1/10<\xi<10$.

\section{Real--world mutualistic-competitive networks}
\label{real}

We now validate the scaling approach to mutualistic-competitive random networks, developed in the 
previous section, by contrasting the obtained universal curves for $\left< \overline{r} \right>$,
$\left< \overline{S} \right>$ and $\left< \overline{R} \right>$ with the spectral and topological properties
of real--world networks. 

To this end we chose a number of pollination networks, host-parasite networks, 
seed dispersal networks and food webs from the Web of Life ecological networks database 
(http://www.web-of-life.es/) with sizes ranging from $n\sim 10$ to $n\sim 1000$; see the adjacency
matrices of some of these mutualistic networks in~\ref{app2}. For each of these networks we 
computed $\left< \overline{r} \right>$, $\left< \overline{S} \right>$ and $\overline{R}$. 
We note that we imposed random weights to the adjacency matrix elements of the real--world 
mutualistic-competitive networks, such that the obtained adjacency matrices are similar to 
those of our RMT model. 
Then, we computed the value of $\xi$ that characterizes each 
of the real--world networks. However, since the real--world networks are highly nonhomogeneous, 
we compute an average sparsity $\left< \alpha \right>$ to be used in $\xi=\left< \alpha \right>n/\lambda$, 
see Eqs.~(\ref{scaling},\ref{xi}). Moreover, notice that $\lambda\equiv\lambda(m/n)$, thus for a given ratio 
$m/n$ we obtain $\lambda$ from $\lambda_r = 2.34 (m/n)^{-0.68}$, 
$\lambda_S = 1.97 (m/n)^{-0.65}$ and $\lambda_R = 0.31 (m/n)^{-1.13}$. Here, the functions 
$\lambda_{r,S,R}$ are power-law fittings to the data, $\lambda$ vs.~$m/n$, reported in Fig.~\ref{Fig9}.
Specifically, $50\%$ of the 160 chosen real-world networks have a $m/n$ ratio in the interval $(1/3,1/2)$; 
while the smallest $m/n$ ratio of our set of real-world networks is about $1/12$.

\begin{figure}[t!]
\centering
\includegraphics[width=0.95\columnwidth]{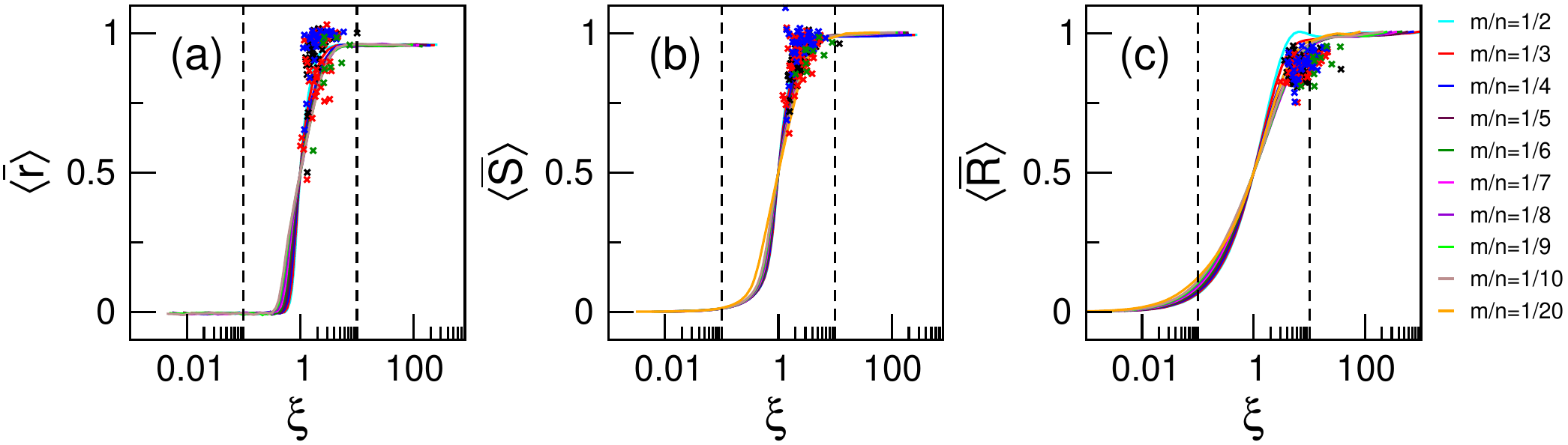}
\caption{Normalized measures (a) $\left< \overline{r} \right>$, (b) $\left< \overline{S} \right>$ and 
(c) $\left< \overline{R} \right>$ as a function of the scaling parameter $\xi$ for random 
mutualistic-competitive networks of size $n=1000$ and several ratios $m/n$. The vertical dashed lines at $\xi=1/10$
and $\xi=10$ mark the onset of eigenvector delocalization and the onset of the GOE regime, respectively.
Crosses indicate the values of (a) $\left< \overline{r} \right>$, (b) $\left< \overline{S} \right>$ and (c) 
$\overline{R}$ of 68 pollination networks (black), 49 host-parasite networks (red), 34 seed dispersal 
networks (blue) and 9 food webs (green) from the Web of Life ecological networks database 
(http://www.web-of-life.es/).}
\label{F10x}
\end{figure}

\begin{figure}[hb!]
\centering
\includegraphics[width=0.95\columnwidth]{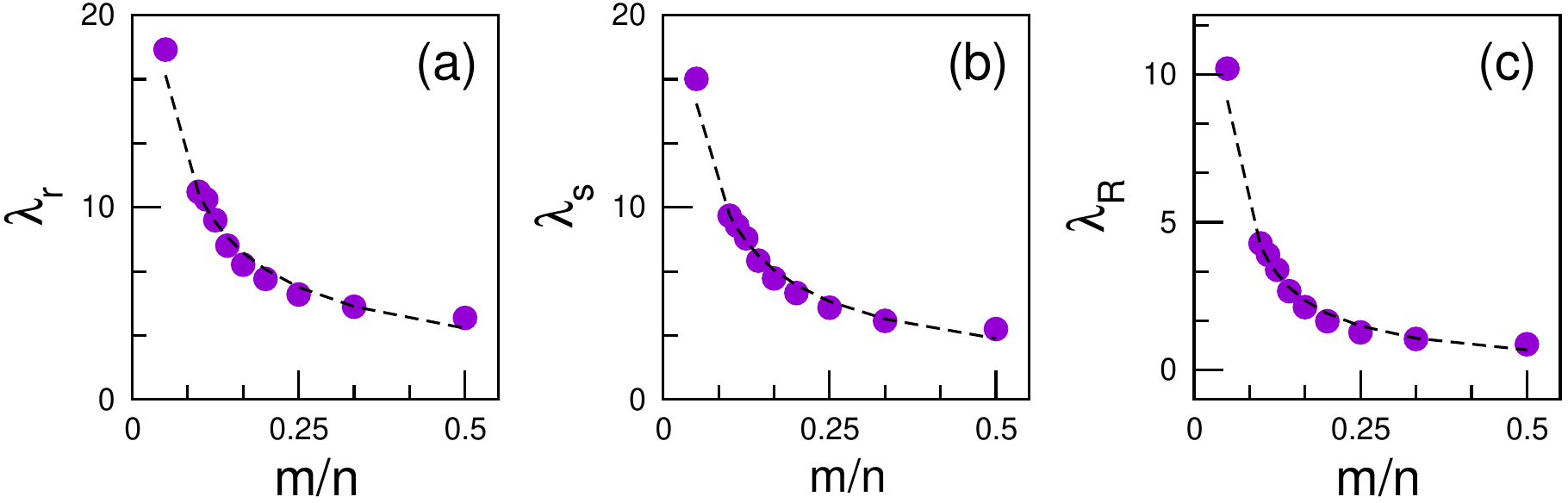}
\caption{Parameter $\lambda$ as function of the ratio $m/n$. $\lambda$ (symbols) was extracted 
from the fittings, with Eq.~(\ref{scaling}), of the curves of Fig.~\ref{Fig5}: (a) $\bra \overline{r} \ket$ 
vs.~$\alpha$, (b) $\bra \overline{S} \ket$ vs.~$\alpha$ and (c) $\bra \overline{R} \ket$ vs.~$\alpha$.
The dashed lines are the best power-law fittings to the data: $\lambda_r = 2.34 (m/n)^{-0.68}$, 
$\lambda_S = 1.97 (m/n)^{-0.65}$ and $\lambda_R = 0.31 (m/n)^{-1.13}$.}
\label{Fig9}
\end{figure}

Finally, in Fig.~\ref{F10x} we report the values of $\left< \overline{r} \right>$, $\left< \overline{S} \right>$ 
and $\overline{R}$ of real--world networks on top of the universal curves obtained from
our RMT approach. Remarkably, we observe a reasonably good correspondence between the spectral 
and topological properties of real--world mutualistic-competitive networks (symbols) and the corresponding statistical
predictions (full lines).

\section{Conclusions}
\label{conclusions}

In this paper, we have applied a statistical approach, based on random matrix theory (RMT) techniques, 
to mutualistic random networks with projected edges that emulate intra-group competitive interactions. 
Specifically, we have proposed a random matrix ensemble that 
represents the adjacency matrices of mutualistic-competitive networks composed by two vertex sets of sizes
$m$ and $n-m$. Thus, the parameters of the RMT model are: the network size $n$, the size of the 
smaller set $m$ (with $1\le m\le n/2$) and the connectivity between the two sets $\alpha\in[0,1]$
forming the mutualistic system.
We focused on the spectral, eigenvector and topological properties of the random network model
by computing, respectively, the ratio of consecutive eigenvalue spacings $r$, the Shannon entropy of the 
eigenvectors $S$ and the Randi\'c index $R$.

First, based on a scaling study, we defined a scaling parameter $\xi\propto \alpha n$, see Eq.~(\ref{xi}), 
that fixes the average spectral, eigenvector and topological properties of the random network model. 
Specifically, we reported universal curves $\left< \overline{X} \right>$ vs.~$\xi$ (where $X$ represents
$r$, $S$ and $R$) that show a weak dependence on the parameter $m$; see Figs.~\ref{Fig7} 
and~\ref{F10x}. Thus, our study provides a way to predict the average properties of random 
mutualistic-competitive networks once $\xi$
is known. On the one hand, concerning the adjacency matrix eigenvectors: For $\xi<1/10$ the 
eigenvectors are localized, $\left< S \right>\approx 0$, when $\xi>10$ the eigenvectors are extended, 
$\left< S \right>\approx \ln(n/2.07)$, whereas the 
localization--to--delocalization transition occurs in the interval $1/10<\xi<10$. Equivalently, 
$\xi \approx 1/10$ marks the onset of eigenvector delocalization (where the adjacency matrix 
eigenvectors cover more than just one vertex in the network), while $\xi \approx 10$ marks 
the onset of the GOE regime (where the adjacency matrix eigenvectors are extended over 
all the vertices forming the network). In this respect, the PE--to--GOE transition reported in 
Sec.~\ref{universality} corresponds to a localization--to--delocalization transition.
On the other hand, concerning the topological properties of the network (that we characterize 
by the use of the Randi\'c index): For $\xi<1/10$ most vertices in the mutualistic-competitive network 
are isolated, $\left< R \right>\approx 0$, while for $\xi>10$ the network acquires the properties 
of a complete network, 
$\left< R \right>\approx n/2$; that is, the transition from isolated vertices to a complete--like 
behaviour occurs in the interval $1/10<\xi<10$. 

Second, we verified our statistical predictions by contrasting them with the properties of real--world 
networks. Indeed, we found a reasonably good correspondence between the properties 
of real--world mutualistic-competitive networks and the corresponding $\left< \overline{X} \right>$ vs.~$\xi$ 
universal curves, as can be clearly seen in Fig.~\ref{F10x}. Of further interest, we 
observed that the real--world networks, even though characterized by values of $\xi$ below the 
onset of the GOE regime, displayed spectral, eigenvector and topological properties very close to 
those of the GOE. This may be understood as a signature of maximal complexity (i.e.~maximal 
chaos in RMT terms) in the real--world mutualistic-competitive networks we analyzed here.


\section*{Acknowledgements}

T.P. acknowledges FAPESP (Grant No. 2016/23827-6).
Y. M. acknowledges partial support from the Government of Aragon, Spain through grant E36-20R (FENOL), by MINECO and FEDER funds (FIS2017-87519-P) and by Intesa Sanpaolo Innovation Center. The funders had no role in study design, data collection, and analysis, decision to publish, or preparation of the manuscript. 

\appendix
\section{Small--network--size effects}
\label{app1}

\begin{figure}[t]
\centering
\includegraphics[width=0.95\columnwidth]{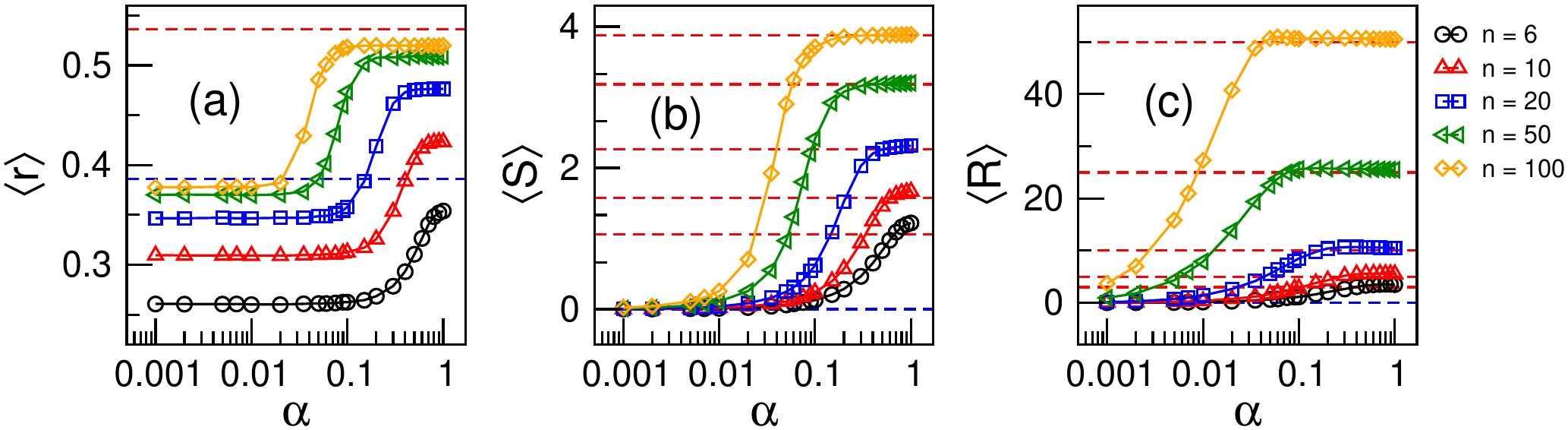}
\caption{(a) Average ratio between consecutive eigenvalue spacings $\bra r \ket$, (b) average 
Shannon entropy $\bra S \ket$ and (c) average Randi\'c index $\bra R \ket$ as a function of the 
connectivity $\alpha$ for random mutualistic-competitive networks of sizes $n\le 100$. Here, $m/n=1/2$ has 
been considered. Each symbol was computed by averaging over $10^6/n$ random networks.
The (blue) red dashed lines in (a) correspond to the RMT predictions for the (PE) GOE.}
\label{FigA10}
\end{figure}

In Fig.~\ref{Fig4} we reported $\bra r \ket$, $\bra S \ket$ and $\bra R \ket$ as a function of the 
connectivity $\alpha$ for mutualistic-competitive random networks of sizes $n\ge 100$ characterized by 
different ratios $m/n$. There, small--network--size effects are evident for $\bra r \ket$; that is
the curves $\bra r \ket$ vs.~$\alpha$ do not approach $\left< r \right>_{\tbox{PE}}\approx 0.3863$ 
and $\left< r \right>_{\tbox{GOE}}\approx 0.5359$ when $\alpha\to 0$ and $\alpha\to 1$, respectively.
As expected, small--network--size effects are even more pronounced for $\bra r \ket$ when $n<100$,
as can be clearly seen in Fig.~\ref{FigA10}(a). For completeness in Figs.~\ref{FigA10}(b) and~\ref{FigA10}(c)
we also present $\bra S \ket$ and $\bra R \ket$, respectively, as a function of the connectivity $\alpha$ 
for mutualistic random networks of sizes $n\le 100$. However, for $\bra S \ket$ and $\bra R \ket$ we 
do not observe important small--network--size effects. In Fig.~\ref{FigA10} we used $m/n=1/2$, but 
other ratios $m/n$ produce similar curves. 

\begin{figure}[t!]
\centering
\includegraphics[width=0.95\columnwidth]{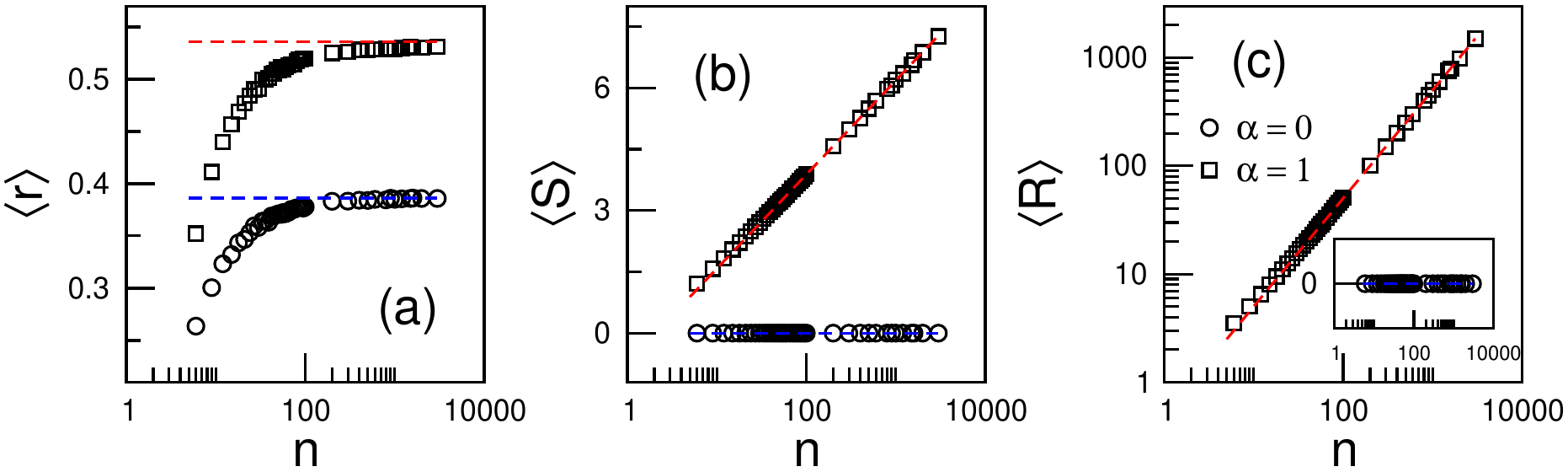}
\caption{(a) $\bra r \ket$, (b) $\bra S \ket$ and (c) $\bra R \ket$ at $\alpha=0$ (circles) and 
$\alpha=1$ (squares) as a function of the graph size $n$. Here, $m/n=1/2$ has been considered. 
The blue [red] dashed lines correspond to the RMT predictions for the PE [GOE]:
$\left< r \right>_{\tbox{PE}}\approx 0.3863$, $\left< S \right>_{\tbox{PE}}=0$ and 
$\left< R \right>_{\tbox{PE}}=0$ [$\left< r \right>_{\tbox{GOE}}\approx 0.5359$, 
$\left< S \right>_{\tbox{GOE}}\approx\ln (n/2.07)$ and $\left< R \right>_{\tbox{GOE}}=n/2$].}
\label{FigA11}
\end{figure}

In particular, as part of the scaling approach to mutualistic-competitive random networks developed in 
Sec.~\ref{universality}, we normalized the spectral and topological measures studied in this paper. 
Specifically, we defined 
$\left< \overline{r} \right>\equiv[\left< r \right>-\left< r \right>_{\tbox{PE}}]/[\left< r \right>_{\tbox{GOE}}-
\left< r \right>_{\tbox{PE}}]$,
$\left< \overline{S} \right>\equiv\left< S \right>/\left< S \right>_{\tbox{GOE}}$ and
$\left< \overline{R} \right>\equiv\left< R \right>/\left< R \right>_{\tbox{GOE}}$.
Therefore, in Fig.~\ref{FigA11} we present $\bra r \ket$, $\bra S \ket$ and $\bra R \ket$ at $\alpha=0$ 
and $\alpha=1$ as a function of $n$ and compare them with the corresponding PE and GOE
predictions, respectively. 
Indeed, since we observe good correspondence between $\bra S \ket$ and $\bra R \ket$ at 
$\alpha=1$ with the corresponding GOE predictions, see Figs.~\ref{FigA11}(b) and~\ref{FigA11}(c),
we used $\left< S \right>_{\tbox{GOE}}\approx\ln (n/2.07)$ and 
$\left< R \right>_{\tbox{GOE}}=n/2$ to compute $\left< \overline{S} \right>$ and
$\left< \overline{R} \right>$, respectively.
In contrast, the RMT predictions for $\left< r \right>$ in the PE and GOE regimes are only approached 
when $n>1000$; see Fig.~\ref{FigA10}(a).
Thus, the values of $\left< r \right>_{\tbox{PE}}$ and $\left< r \right>_{\tbox{GOE}}$ used
to compute $\left< \overline{r} \right>$ in Sec.~\ref{universality} were calculated numerically for
the given network sizes used.

\section{Adjacency matrices of real--world networks}
\label{app2}

\begin{figure}
\centering
\includegraphics[width=0.99\columnwidth]{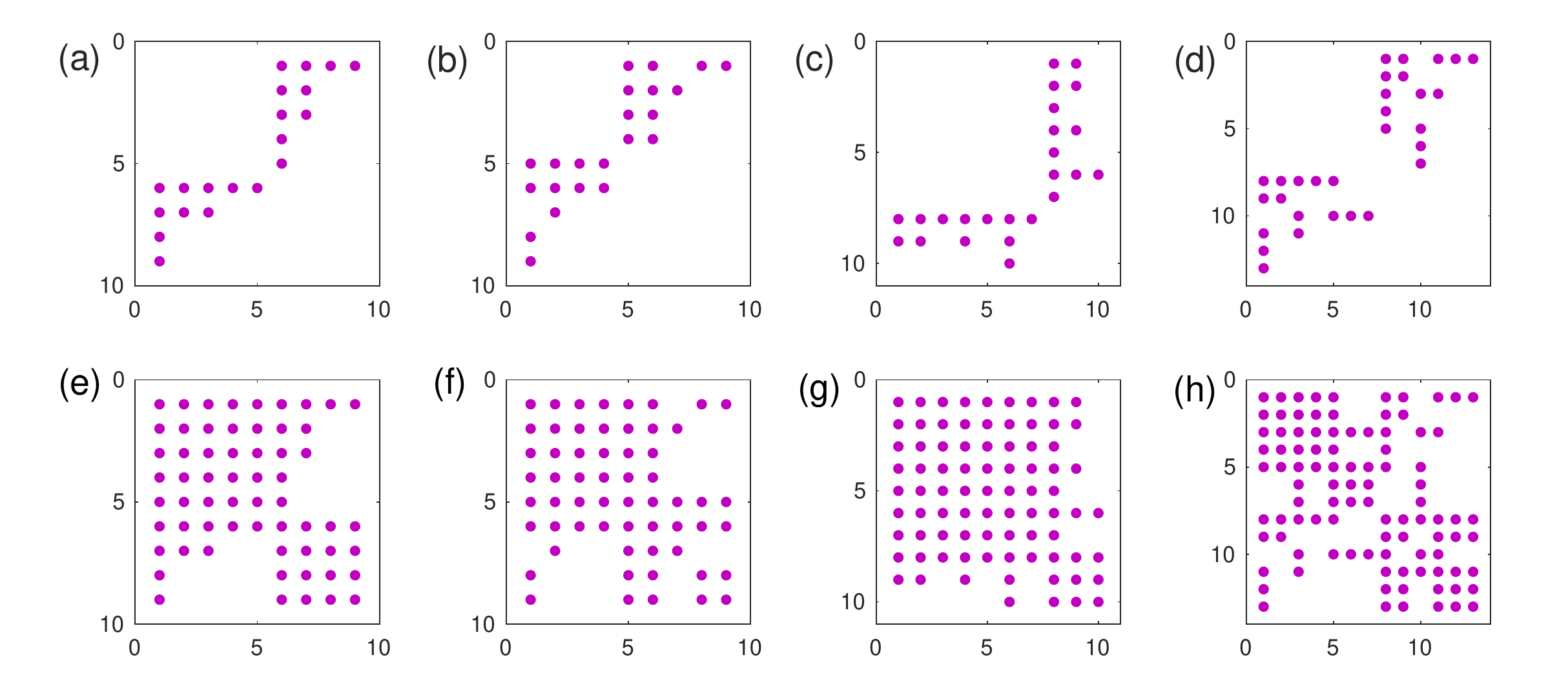}
\caption{Top panels: Adjacency matrices of real--world bipartite networks from the Web of 
Life ecological networks database (http://www.web-of-life.es/). 
(a) Seed dispersal network $\mbox{M\_SD\_029}$ ($n=9$, $m=4$, $\alpha=0.5$), 
(b) seed dispersal network $\mbox{M\_SD\_030}$ ($n=9$, $m=5$, $\alpha=0.55$), 
(c) host parasite network $\mbox{A\_HP\_015}$ ($n=10$, $m=3$, $\alpha=0.571$) and 
(d) host parasite network $\mbox{A\_HP\_035}$ ($n=13$, $m=6$, $\alpha=0.357$). 
Lower panels: Corresponding mutualistic-competitive adjacency matrices.}
\label{FigB12}
\end{figure}

\begin{figure}
\centering
\includegraphics[width=0.99\columnwidth]{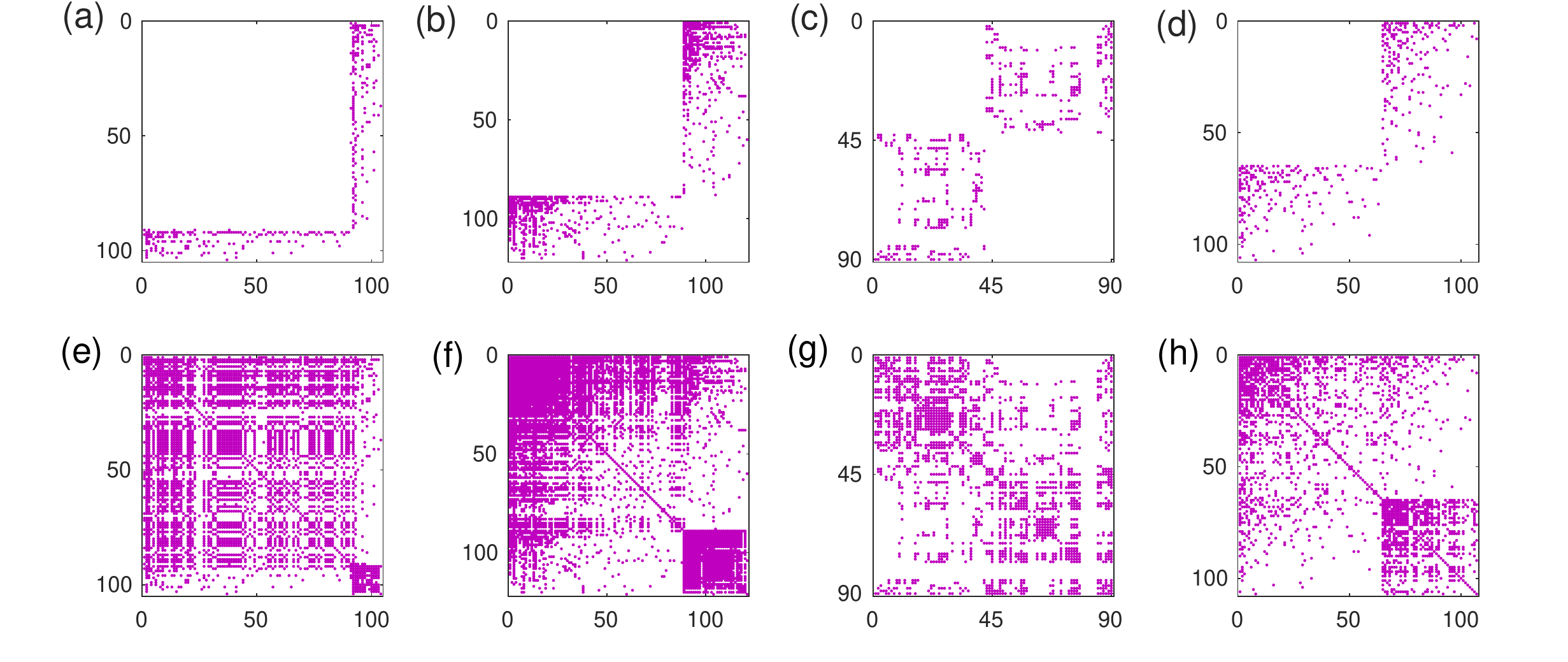}
\caption{Top panels: Adjacency matrices of real--world bipartite networks from the Web of 
Life ecological networks database (http://www.web-of-life.es/). 
(a) Pollination network $\mbox{M\_PL\_051}$ ($n=104$, $m=14$, $\alpha=0.13$), 
(b) seed dispersal network $\mbox{M\_SD\_034}$ ($n=121$, $m=33$, $\alpha=0.139$, 
(c) food web network $\mbox{FW\_007}$($n=90$, $m=42$, $\alpha=0.109$) and 
(d) pollination network $\mbox{M\_PL\_002}$ ($n=107$, $m=43$, $\alpha=0.071$). 
Lower panels: Corresponding mutualistic-competitive adjacency matrices.}
\label{FigB13}
\end{figure}

\begin{figure}
\centering
\includegraphics[width=0.99\columnwidth]{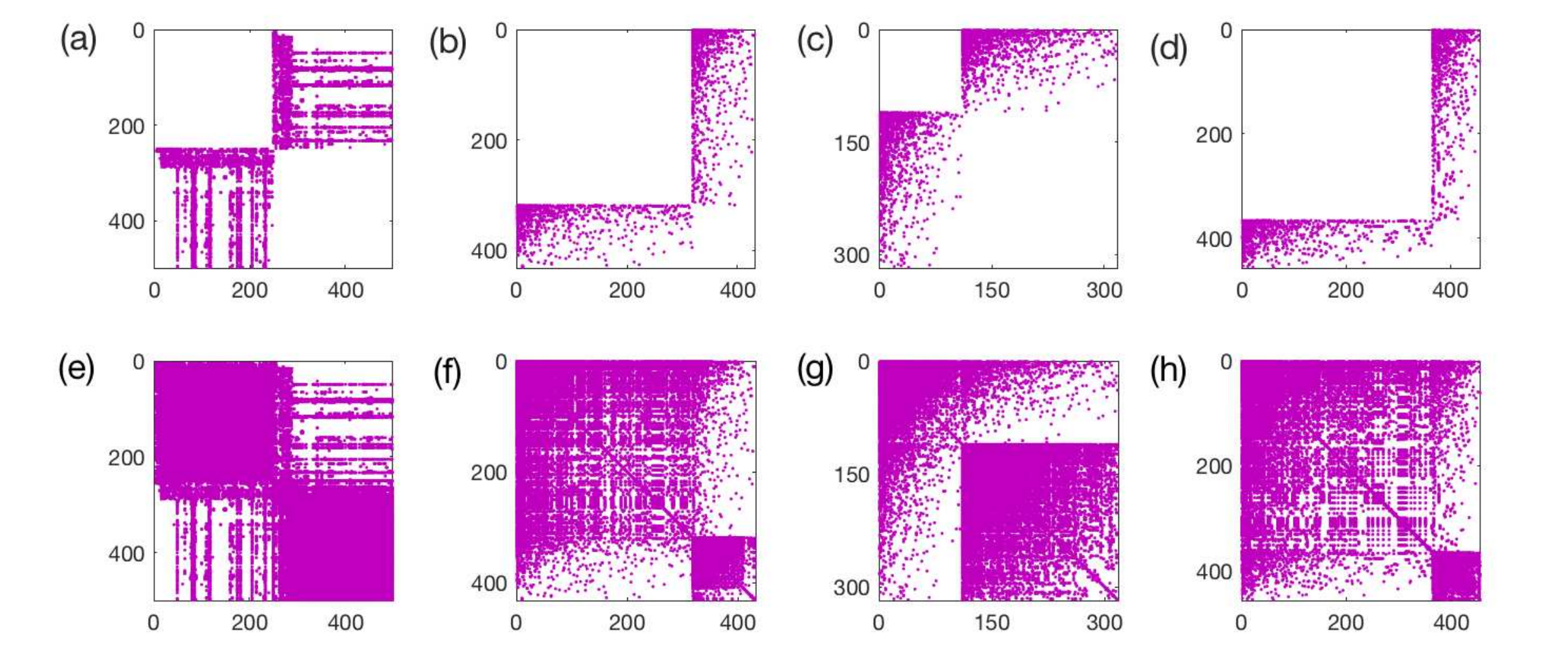}
\caption{Top panels: Adjacency matrices of real--world bipartite networks from the Web of 
Life ecological networks database (http://www.web-of-life.es/).  
(a) Food web network $\mbox{FW\_008}$($n=498$, $m=249$, $\alpha=0.0534$), 
(b) pollination network $\mbox{M\_PL\_054}$ ($n=431$, $m=113$, $\alpha=0.022$), 
(c) seed dispersal network $\mbox{M\_SD\_022}$ ($n=317$, $m=110$, $\alpha=0.049$ and 
(d) pollination network $\mbox{M\_PL\_056}$ ($n=456$, $m=91$, $\alpha=0.026$). 
Lower panels: Corresponding mutualistic-competitive adjacency matrices.}
\label{FigB14}
\end{figure}

In Sec.~\ref{real} we validated the scaling approach to mutualistic-competitive random networks developed in 
Sec.~\ref{universality} by contrasting the obtained universal curves for $\left< \overline{r} \right>$,
$\left< \overline{S} \right>$ and $\left< \overline{R} \right>$ with the spectral and topological properties
of real--world networks. Here, in Figs.~\ref{FigB12}, \ref{FigB13} and~\ref{FigB14} we present the
actual adjacency matrices of some of the real--world networks from the Web of Life ecological 
networks database (http://www.web-of-life.es/).
In Figs.~\ref{FigB12}, \ref{FigB13} and~\ref{FigB14} we report some examples of small--size ($n\sim 10$), 
medium--size ($n\sim 100$) and large--size ($n\sim 500$) networks, respectively.
For completeness, in each figure we show bipartite adjacency matrices (upper panels) as well as the 
corresponding mutualistic-competitive adjacency matrices (lower panels).




\end{document}